# Towards in vivo photon dosimetry using radiation-sensitive mixed-core ultrasound phase-change contrast agents

Bram Carlier[1,2], Reza Bahrami Gorji[2], Qingyuan Tan[3], Meritxell Gambús-Marfà[1,4], Uwe Himmelreich[1], Meng-Xing Tang[3], Koen Van Den Abeele[2], Sophie V. Heymans[2,5]

[1]Department of Imaging & Pathology, KU Leuven, Leuven, Belgium
[2]Department of Physics, KU Leuven, Kortrijk, Belgium
[3]Department of Bioengineering, Imperial College London, United Kingdom
[4]Institut Químic de Sarrià (IQS), Universitat Ramon Llull (URL), 08017, Barcelona, Spain
[5]Department of Cardiovascular Sciences, KU Leuven, Leuven, Belgium

## 0. Abstract

*Objective*. Advanced radiotherapy treatments apply complex treatment plans that require careful verification of the beam delivery. Therefore, in vivo dosimetry, measuring delivered radiation doses directly on the patient, has become increasingly important. Radiation-sensitive phase-change nanodroplets were recently proposed for in vivo radiation sensing. The radiation-induced conversion of superheated nanodroplets into echogenic microbubbles generates ultrasound contrast from which radiation deposition can be derived. While successful for in vivo proton range verification, previously used nanodroplet formulations were not suitable for in vivo photon dosimetry due to stringent requirements on the degree of superheat of the nanodroplet core. Here, we used a mixed-core strategy, combining different perfluorocarbons in the same nanodroplet core, to modulate the photon response and droplet stability at body temperature.

*Approach*. Lipid-shelled nanodroplets with different cores (pure octafluoropropane, pure decafluorobutane and a 17:4 (v/v) octafluoropropane:decafluorobutane mixture) were dispersed in deionized water in phantom containers and irradiated with a preclinical photon irradiator at various temperatures. Simultaneous high-frame rate ultrasound acquisition allowed to link ultrasound contrast generation to the radiation delivery. Nanodroplet yield and composition were determined using $^{19}F$ NMR spectroscopy.

*Main results*. For the first time, photons induced nanodroplet-mediated ultrasound contrast strong enough for clinical applications at body temperature. A clear trade-off was observed between a stronger photon response (increased degree of superheat) and maintaining sufficient droplet stability. Neither octafluoropropane (lacking stability) nor decafluorobutane (lacking photon response) cores proved suitable. On the contrary, the mixed-core formulation exhibited a strong photon response with minimal background already at small radiation doses (0.1 Gy). Future work will optimize the nanodroplet synthesis to improve reproducibility in view of quantitative dosimetric read-outs.

*Significance*. We demonstrated the feasibility to modulate the photon radiation response of superheated nanodroplets at body temperature using a mixed-core formulation. This yielded radiation-induced ultrasound contrast sufficiently strong for future in vivo dosimetry.

## 1. Introduction

Radiotherapy is one of the cornerstones of cancer therapy, able to benefit about half of the cancer patients (Lievens *et al* 2020). Technological advances facilitate highly conform treatment delivery, but the associated steep dose gradients make treatments prone to uncertainties. These arise from the image-based treatment planning process, patient motion, setup errors or anatomical changes throughout the treatment, and can result in underdosage of the tumor or excessive healthy tissue exposure if not properly accounted for (Mijnheer *et al* 2013).

In vivo dosimetry (IVD) measures the received radiation dose directly on or inside the patient (Mijnheer *et al* 2013), ensuring correct dose delivery to the tumor target, preventing erroneous healthy tissue exposure and contributing to quality assurance protocols (Chow and Ruda 2025). As a result, IVD has become increasingly important in the radiotherapy workflow, with a plethora of dosimetry solutions available. One-dimensional dosimeters including thermoluminescent detectors (TLDs), optically-stimulated luminescent detectors (OSLDs), plastic scintillating detectors (PSDs), diodes, and metal oxide semiconductor field-effect transistors (MOSFETs) have a strong clinical track record, but are essentially point based measurement devices and thus not suitable to capture complex 3D dose distributions (Olaciregui-Ruiz *et al* 2020). While the use of such detectors in array configurations solves part of the problem, the accurate evaluation of sharp dose gradients still relies strongly on the detector positioning, which is typically of low spatial resolution and limited to the patient's surface. Film dosimeters provide highly resolved 2D dose information, but require intensive processing (radiographic) or strict read-out protocols (radiochromic) (Bjegovic *et al* 2025, Olaciregui-Ruiz *et al* 2020, Mijnheer *et al* 2013). Moreover, they are made for single use and positioning is mostly limited to the surface of the patient. Electronic Portal Imaging Devices (EPIDs) use flat panel detectors integrated in the treatment device to monitor the transit dose. The dose delivery inside the patient is determined through the use of patient CT data and back projection algorithms (Mijnheer 2017). While relatively easy to integrate in current radiotherapy workflows, with several commercial solutions available (Kadoya *et al* 2018, Olaciregui-Ruiz *et al* 2020), some limitations remain. For example, the quality of the dose information depends on the adequacy of the used CT scans and the lack of tissue-equivalence of the detector requires careful calibration to perform absolute dosimetry (Nelms *et al* 2010). In addition, EPID dosimetry is restricted to applications where a clear transit dose is present, making it unsuitable for hadron therapy or internal radiotherapy (Bjegovic *et al* 2025). Cherenkov imaging can be used to verify dose delivery by capturing the visual photons released when charged particles move faster through tissue than the speed of light (Darafsheh *et al* 2024). Given the limited penetration depth of visual light, Cherenkov imaging is however limited to superficial applications such as skin and breast irradiation. Finally, radiacoustic imaging utilizes ultrasound waves generated during the thermoelastic expansions of tissues upon pulsed dose delivery to capture true 3D dose distributions directly in the patient (Sun *et al* 2023, Bjegovic *et al* 2025). While particularly promising for FLASH radiotherapy, where high radiation doses are delivered in short pulses, the technique is still in its infancy with many challenges remaining, e.g. poor signal-to-noise ratio (SNR), limited view artifacts, and the complex influence of tissue heterogeneity and acoustic attenuation on the quantitative relationship between the detected ultrasound signals and the radiation dose (Bjegovic *et al* 2025).

To overcome the limitations of currently available in vivo dosimetry solutions, we developed radiation-sensitive phase changing ultrasound contrast agents. In particular, superheated nanodroplets can convert into echogenic microbubbles upon irradiation, thereby generating strong ultrasound contrast from which the dose delivery can be assessed (Carlier *et al* 2020, Heymans *et al* 2021). Nanodroplets can be sensitized to different types of ionizing radiation by varying the degree of superheat of their liquid core as described by the theory of radiation-induced nucleation of superheated emulsions (D'Errico 2001, Heymans *et al* 2021). The reduced degree of superheat ($s$, equation 1) is a key parameter to characterize the radiation sensitivity:

$$s = \frac{T - T_b}{T_c - T_b} \qquad (eq. 1)$$

with $T_b$ and $T_c$ referring to the boiling and critical temperature of the superheated liquid, and $T$ the operation temperature. With increasing degree of superheat, less energy is required to trigger droplet vaporization, hence ionizing radiation with lower linear energy transfer (LET) can initiate droplet vaporization. Specifically, protons can induce ultrasound contrast from $s \approx 0.42$, while electrons released during photon therapy require at least $s \approx 0.51$ (D'Errico 2001), as their peak LET is lower. Above the sensitization limit, the radiation response increases exponentially, until $s \approx 0.65$, beyond which the liquid becomes so unstable that spontaneous vaporization occurs. Note that these values were determined for naked microemulsions, while the nanodroplets used in our previous and current studies are additionally encapsulated by a polymeric or lipidic shell.

We confirmed the potential of radiation-sensitive nanodroplets for in vivo proton range verification using nanodroplets with a perfluorobutane core and a poly(vinyl alcohol) shell (Carlier *et al* 2024a, Collado-Lara *et al* 2022). However, for photon radiotherapy, the low LET of electrons required excessive heating of the perfluorobutane core (to ~65 °C) before a photon response was observed, preventing in vivo applications (Carlier *et al* 2024b). To achieve the same degree of superheat and thus radiation response at body temperature, a nanodroplet core of lower boiling point is needed. At the same time, it is important that the nanodroplets do not become too unstable to handle.

In the context of acoustic droplet vaporization (ADV) of phase-change contrast agents, where droplet vaporization is triggered by high-pressure ultrasound waves instead of ionizing radiation, a similar trade-off was observed in terms of droplet stability and acoustic pressures required to trigger phase change. This was resolved by combining multiple perfluorocarbon liquids in the same nanodroplet core, allowing to tune the required vaporization pressure by changing the ratio of the core liquids (Matsunaga *et al* 2012, Riemer *et al* 2023).

Here, we utilized this mixed-core approach for the first time to tune the nanodroplet radiation sensitivity to photons at body temperature (37 °C). First, we illustrate the influence of the degree of superheat of the nanodroplet core on the photon response and demonstrate the inadequacy of pure perfluorocarbon cores. On the contrary, a 17:4 (v/v) OFP:DFB mixed-core combines a good stability at body temperature with a strong radiation response. For this formulation, we evaluated the reproducibility of the photon response under different irradiation conditions and linked our observations back to the nanodroplet synthesis. Finally, we explored the necessary next steps towards quantitative dosimetry.

## 2. Materials & Methods

### 2.1. Nanodroplet preparation via the microbubble condensation method

Radiation-sensitive nanodroplets were prepared from microbubble precursors as described previously (Zhang *et al* 2019, Matsunaga *et al* 2012). Briefly, a lipid mixture was prepared by dissolving 1,2-distearoyl-sn-glycero-3-phosphocholine (DSPC) and 1,2-distearoyl-sn-glycero-3-phosphoethanolamine-N-[amino(polyethylene glycol)-2000] (DSPE-PEG-2000) in a 9:1 molar ratio. The solvent used was a 16:3:1 (v/v) solution of phosphate-buffered saline (PBS), propylene glycol, and glycerol. Lipids were purchased from Avanti Polar Lipids, Inc. (Alabaster, AL, USA), and solvents were obtained from Merck (Hoeilaart, Belgium). A volume of 1 mL of the resulting solution with 1 mg/mL total lipid concentration was transferred to a 2 mL glass vial and sealed hermetically. Afterwards, 21 mL of perfluorocarbon gas was injected in the headspace (~17 mL with relief needle and the final ~4 mL after removal of the relief needle to allow pressure build-up). The perfluorocarbon gas was either pure OFP, pure DFB or a 17:4 (v/v) mixture OFP:DFB. After injection, the glass vial was mechanically agitated for 3 x 30 s in a bubble shaker (Capmix, 3M ESPE Deutschland Gmbh, Seefeld, Germany). This microbubble precursor solution was consequently submerged in an ice-salt bath (~ -5 °C) for 5-10 minutes. Then, the solution was transferred to a 1 mL Luer Lock syringe, subsequently sealed with a closed stopcock valve and eventually manually pressurized until the opaque bubble solution turned transparent, indicating successful microbubble condensation. Droplets were stored on ice and used immediately (<30 minutes) after preparation.

### 2.2. Nanodroplet / microbubble characterization

The nanodroplet yield and core composition were determined by quantifying the perfluorocarbon content using $^{19}$F NMR spectroscopy (400 MHz Avance II NMR spectrometer, Bruker Biospin GmbH, Rheinstetten, Germany). We added an internal standard of known concentration (200 µL of 20 mM fluorocytosine) to the 5 mm NMR tube containing 100 µL of nanodroplets and 100 µL deuterated water. We used a free induction decay size of 65536 data points, and acquired a spectral width of 249 ppm with a transmitter frequency of 376.49 MHz. Signals were averaged 880 times, yielding an acquisition time of 20 minutes. To quantify the NMR spectra, peaks were integrated after phase and baseline correction using Topspin Processing 5.0.0. Size distribution and number concentration of the microbubble precursors were characterized via Coulter counter measurements (Multisizer 4e, Beckman Coulter, Inc., Brea, CA, USA) using a 20 µm aperture tube. For the measurements, a sample of 50 µL was added to 100 mL filtered Isoton II diluent and continuously stirred at 5 rpm.

### 2.3. Experimental setup

The photon radiation response of the different nanodroplet formulations was evaluated using a Small Animal Radiation Research Platform (SARRP) from Xstrahl Ltd. (Walsall Wood, United Kingdom), featuring a 225 kV X-ray tube. Nanodroplets were homogenously dispersed in 100 mL deionized water in custom, 3D-printed phantom containers (6 x 5 cm$^2$, figure 1). The latter contained imaging fiducials (purple arrows in figure 1b) visible in cone-beam computed tomography (CBCT) to facilitate a reproducible positioning of the isocenter. A single, 10 x 10 mm$^2$ radiation field was applied using a brass collimator and placed within the left half of the phantom container as seen in the yellow pane (bottom right) in figure 1b. This way, the right half of the phantom served as an internal control, albeit still subject to stray radiation. A thermocouple was placed inside the phantom container to monitor the temperature and to provide feedback to control an integrated heating element maintaining the temperature during long (2 Gy) irradiations.

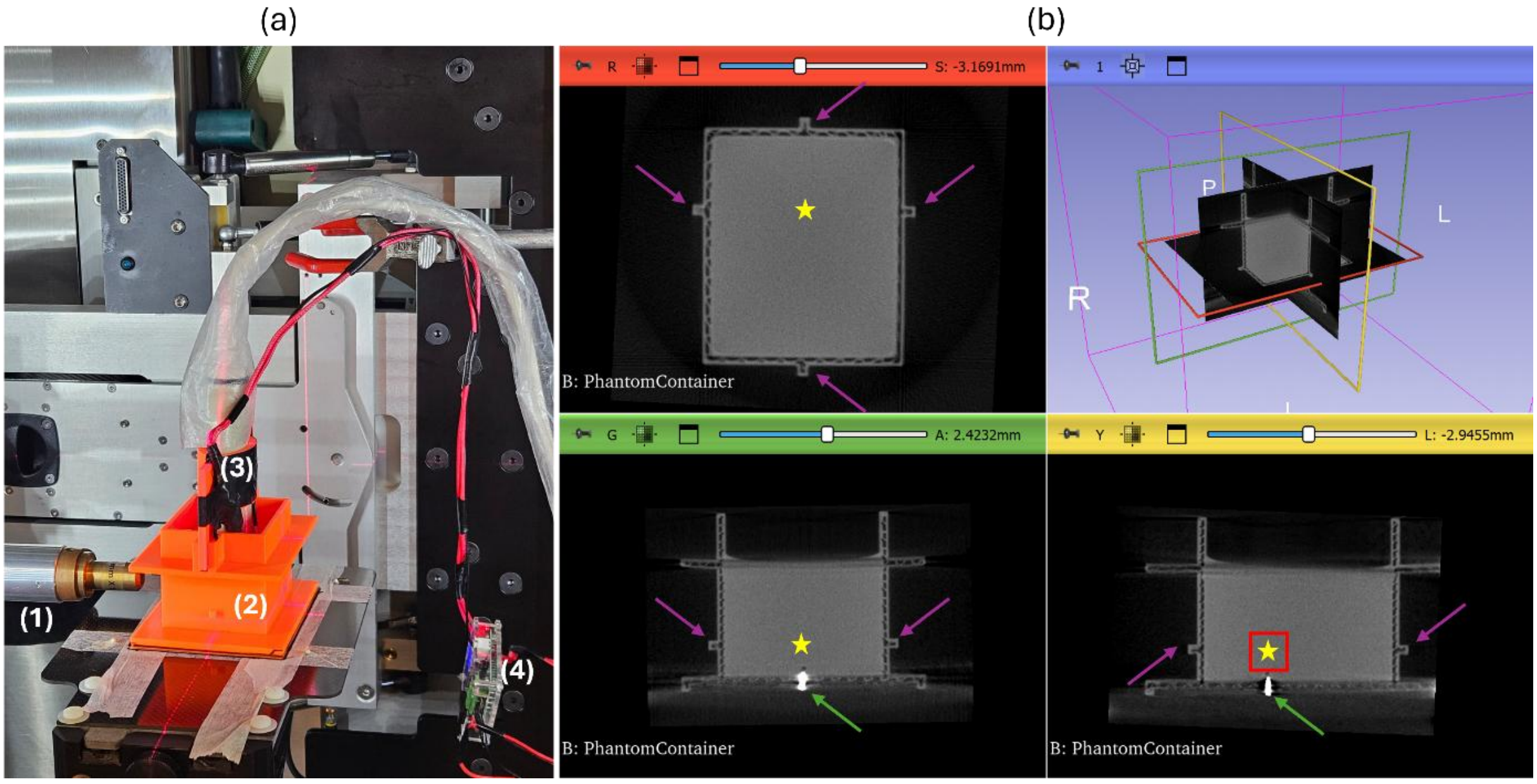

**Figure 1.** Experimental set-up of the phantom irradiation experiments. (a) Photograph of the experimental set-up indicating its most important elements: (1) photon beam shaped to a 10 x 10 $mm^2$ radiation field using a brass collimator; (2) 3D-printed phantom container containing freely floating nanodroplet samples in deionized water; (3) ultrasound probe in a probe holder connected to the phantom container for reproducible ultrasound imaging during irradiation; (4) temperature control unit consisting of a thermocouple placed in the water in the phantom, a temperature display and a heating element to maintain the temperature during the long (2 Gy) irradiations. (b) CBCT scan of the water-filled phantom container acquired by the SARRP system and used to design the radiation treatment plan. Purple and green arrows indicate imaging fiducials added to the phantom container to reproducibly position the isocenter (yellow star). The yellow pane (bottom right) aligns with the plane of ultrasound imaging and the red box represents the 10 x 10 $mm^2$ radiation field.

### 2.4. Ultrasound imaging

A 7.5 MHz L7-XTech probe (Vermon, Tours, France) was positioned onto the phantom container using a probe holder for online ultrasound imaging during irradiation. The probe was connected to an ULA-OP 256 research platform (X-Phase srl, Florence, Italy), which was located just outside of the shielded SARRP system. Two different imaging sequences were used depending on the frame rate and scan duration. For long acquisitions, all 128 transducer elements of the probe were used to transmit plane waves (15 angles compounding from -7.5 to +7.5 degrees), with a pulse repetition frequency of 400 Hz, resulting in a frame rate of 27 Hz. Given the available buffer size, this allowed for a total acquisition time of ~100 s, able to capture the full delivery of a 2 Gy fraction (~72 s). For high-frame rate imaging, a pulse repetition frequency of 6 kHz resulted in a frame rate of 400 Hz, able to capture the fast response dynamics at the onset of irradiation in a ~7 s acquisition window. Again, 15 plane waves were compounded as before. Depending on the nanodroplet core and operation temperature, different (low) transmit acoustic pressures were used to ensure no vaporization events were triggered acoustically.

### 2.5. In phantom photon radiation response

The radiation conditions for the different phantom experiments are listed in table 1. The response of pure OFP droplets was evaluated at increasing temperatures: 21 °C, 28 °C and 33 °C. For the mixed-core droplets, all irradiations were performed at 37 °C. Most irradiations were imaged using the high-frame rate sequence. Given the limited buffer size, radiation dose was set to 0.1 Gy, which corresponded to an irradiation of ~4 s at the maximally achievable dose rate of 1.67 Gy/min. A dose of 2 Gy was applied for the low-frame rate acquisition. Finally, two negative control experiments were conducted: the irradiation of (i) one phantom without droplets and (ii) three phantoms with pure DFB droplets, featuring a reduced degree of superheat below the expected photon sensitization limit.

**Table 1.** Phantom irradiation conditions.

| Core composition | Operation temperature* | s | Droplet volume [µL] | Number of measurements† | Dose [Gy] | US imaging duration [s] | Frame rate [Hz] |
|---|---|---|---|---|---|---|---|
| / | 37 °C | / | 0 | 1 | 0.1 | ~7 | 400 |
| Pure OFP | 21 °C | 0.53 | 200 | 6 (2 x 3) | 0.1 | ~7 | 400 |
| | 28 °C | 0.60 | 200 | 5 (3 + 1 + 1) | 0.1 | | |
| | 33 °C | 0.64 | 200 | 3 | 0.1 | | |
| 17:4 OFP:DFB | 37 °C | ≈0.63‡ | 200 | 12 (4 x 3) | 0.1 | ~7 | 400 |
| | | | 50 | 6 (2 x 3) | 2 | ~100 | 27 |
| | | | 200 | 6 (2 x 3) | 2 | ~100 | 27 |
| Pure DFB | 37 °C | 0.34 | 200 | 3 | 0.1 | ~7 | 400 |

*While slight fluctuations of the temperature occurred during the different experiments, the deviation from the target temperature was kept ≤1 °C unless indicated differently.

†Numbers in between brackets reflect the amount of repeats per nanodroplet vial, e.g. (2 x 3) means three repeats were performed for each of the two independent nanodroplet vials.

‡Calculated using Raoult’s law, assuming the core behaves as an ideal mixture of liquids.

### 2.6. Image processing

Raw ultrasound signals were first beamformed using the delay-and-sum method (Perrot *et al* 2021) weighted by the coherence factor (Matrone *et al* 2020), with subsequent coherent compounding of the 15 transmitted plane waves. Then, the obtained images were bandpass filtered (Butterworth, 2-20 MHz), envelope-detected through Hilbert transformation and eventually log-compressed to dB-scale. Finally, gray values across all images were normalized by subtraction of a fixed value (75 dB). Average gray values throughout the acquisition were determined in a 10 x 10 $mm^2$ region-of-interest (ROI) delimiting the radiation field and positioned using the imaging fiducial visible both in CBCT and ultrasound images (green arrow in figure 1b). As saving of the ultrasound data was triggered manually, small variations in the onset of the radiation response during the ultrasound acquisition were expected. Therefore, the start of irradiation was determined from the average gray value in the ROI as follows. For high frame rate acquisitions, first a 100 ms moving average was applied to the quantified response. Then, the first derivative was calculated, which was subsequently filtered by a 100 ms (40 frames) moving average. The frame of radiation onset was then defined as the first of hundred consecutive frames with a positive slope exceeding an empirical threshold. For low frame rate acquisitions, the same approach was followed, but the moving average filter was changed to 111 ms (3 frames) and the onset frame was the first of 27 consecutive frames with a positive slope exceeding the threshold. In cases where this approach failed due to a low radiation response, either the complete ultrasound acquisition is shown or the start of irradiation was set to the theoretical start (after 3 s of the 7 s acquisition). For visualization purposes, displayed quantifications were cropped from 1 s before the presumed irradiation onset to the end of the irradiation.

## 3. Results

### 3.1 The degree of superheat of the nanodroplet core strongly influences the photon response

To illustrate the importance of the degree of superheat of the nanodroplet core, we first irradiated pure OFP droplets at different temperatures covering the complete photon-sensitive zone: 21 °C (s = 0.53), 28 °C (s = 0.60) and 33 °C (s = 0.64). The photon response and its dynamics are displayed in figure 2.

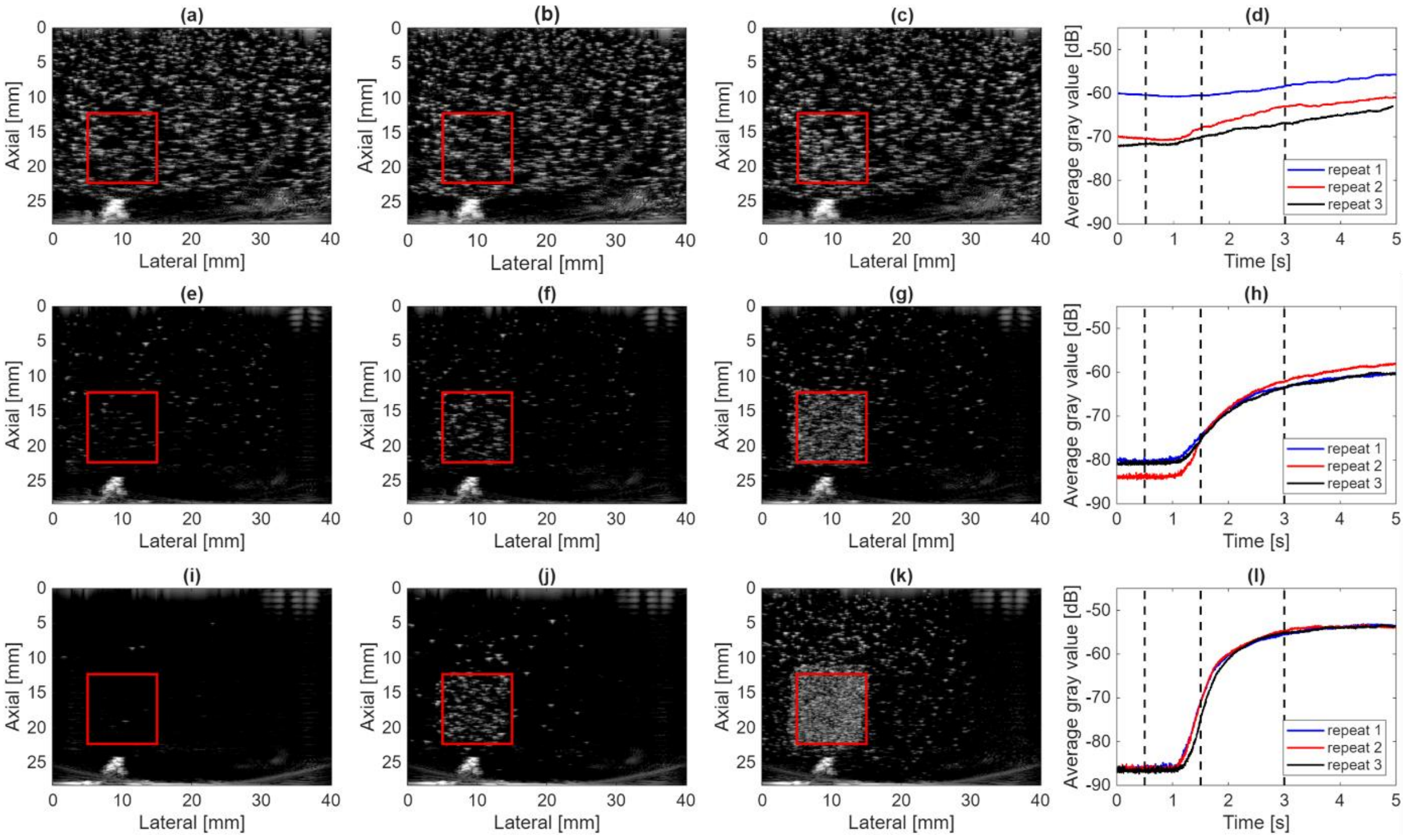

**Figure 2.** Photon radiation response of pure OFP nanodroplets at different degrees of superheat. B-mode images at different time points with respect to the start of the irradiation: -0.5 s (a, e, i), +0.5 s (b, f, j) and +2 s (c, g, k), shown for a representative phantom at each temperature: 21 °C (a-c), 28 °C (e-g) and 33 °C (i-k). The phantoms were irradiated with a dose of 0.1 Gy at a dose rate of 1.67 Gy/min. The red box indicates the 10 x 10 $mm^2$ radiation field and quantification ROI. The final column (d, h, l) shows the corresponding dynamics of the quantification of the average gray value for the entire ultrasound acquisition for three different phantoms (prepared from the same vial). The red curve corresponds to the displayed B-mode images. Vertical dashed lines correspond to the different time points of the B-mode images (irradiation starts at ~1 s).

At 21 °C, just above the photon sensitization threshold, we notice that the radiation response is small with a near linear increase of the average gray value after irradiation started (figure 2d). Note the large amount of background bubbles. As these were already present before irradiation started, they are attributed to suboptimal nanodroplet synthesis (relying on microbubble condensation). Indeed, a repeat of the radiation experiment at 21 °C with a new nanodroplet vial barely showed any background signals (figure S1). At 28 °C, a much stronger radiation response was observed with barely any background bubbles prior to irradiation. Initially, gray value increase is near linear, but rapidly slows down towards a plateau. Finally, at 33 °C, the radiation-induced ultrasound contrast follows a clear sigmoid curve, with a fast rise time. As droplets and radiation-induced microbubbles were free-floating, the microbubble cloud flows out of the radiation field over time. Sparse vaporizations appear beyond the red radiation field as a result of stray radiation. We also attempted to irradiate the pure OFP nanodroplets at 37 °C (s = 0.68). However, a high degree of spontaneous vaporization at this elevated degree of superheat prevented an ultrasound read-out.

Despite variability in the background levels as reflected in the gray value level before irradiation starts, the dynamics of the contrast generation were highly reproducible for different phantoms prepared from the same nanodroplet vial, indicating high intra-vial reproducibility.

Finally, to confirm ultrasound contrast was solely induced by radiation-induced droplet vaporization, we also irradiated a water phantom without droplets at 37 °C. As expected, no microbubbles were formed and hence no radiation-induced ultrasound contrast was generated (figure S2).

### 3.2 DFB nanodroplets are insufficiently superheated to generate a photon response at body temperature

To overcome the poor stability of OFP nanodroplets at body temperature, higher boiling point liquid cores can be used. For instance, we repeated the previous radiation experiments at 37°C with pure DFB droplets (figure 3). Due to the boiling point of -2 °C, DFB droplets feature a reduced superheat of only s = 0.34 at 37 °C. This explains the lack of radiation response observed in all of the irradiated phantoms.

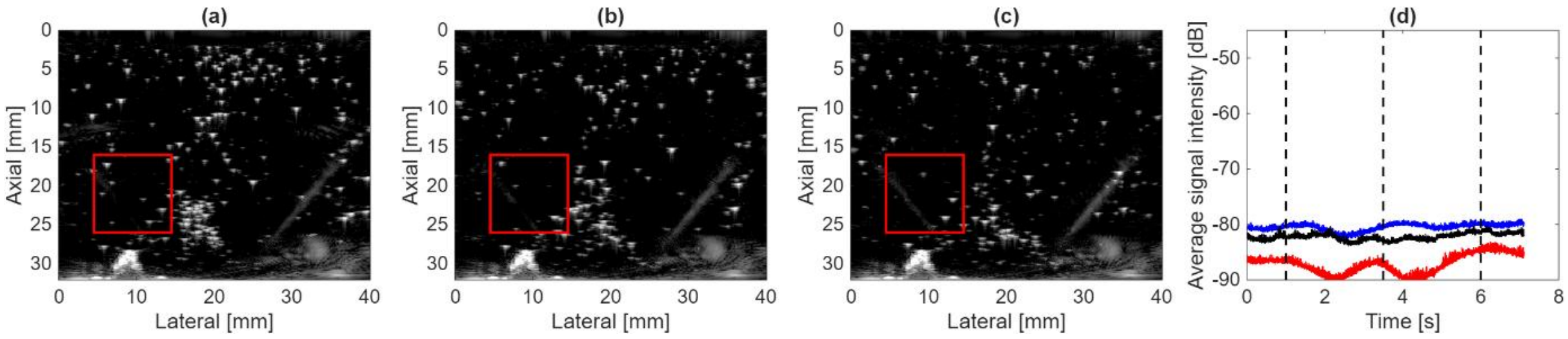

**Figure 3.** Photon radiation response of pure DFB nanodroplets at 37 °C. B-mode images at different time points during ultrasound acquisition (1 s, 3.5 s, 6 s) are shown for a representative phantom (a-c). The phantom was irradiated with a dose of 0.1 Gy at a dose rate of 1.67 Gy/min. The irradiation starts at ~3 s. The red box indicates the 10 x 10 $mm^2$ radiation field and quantification ROI. The final column (d) shows the corresponding dynamics of the quantification of the average gray value for the entire ultrasound acquisition for three different phantoms (prepared from the same vial). The red curve corresponds to the displayed B-mode images. Vertical dashed lines correspond to the different time points of the B-mode images. Note that for the black curve in panel (d), we only reached 35.5 °C during irradiation.

These observations highlight that pure liquid cores lack the flexibility to tune the nanodroplet degree of superheat to exhibit a strong photon response, while maintaining good stability at body temperature.

### 3.3 Mixed-core nanodroplets allow to tune the degree of superheat to generate a strong radiation response at body temperature

To overcome the limitations of pure PFC cores, we propose to mix different perfluorocarbons to obtain a photon-sensitive nanodroplet formulation at body temperature. Theoretical calculations using Raoult's law estimate that for a 17:4 volumetric ratio of OFP:DFB the reduced degree of superheat is $s \approx 0.63$, assuming an ideal mixture behavior. Figure 4 shows the photon response of these mixed-core nanodroplets after irradiation at 37 °C for 12 different irradiations, spread over four different nanodroplet vials. A video of the radiation response can be found in the supplementary materials (video S1).

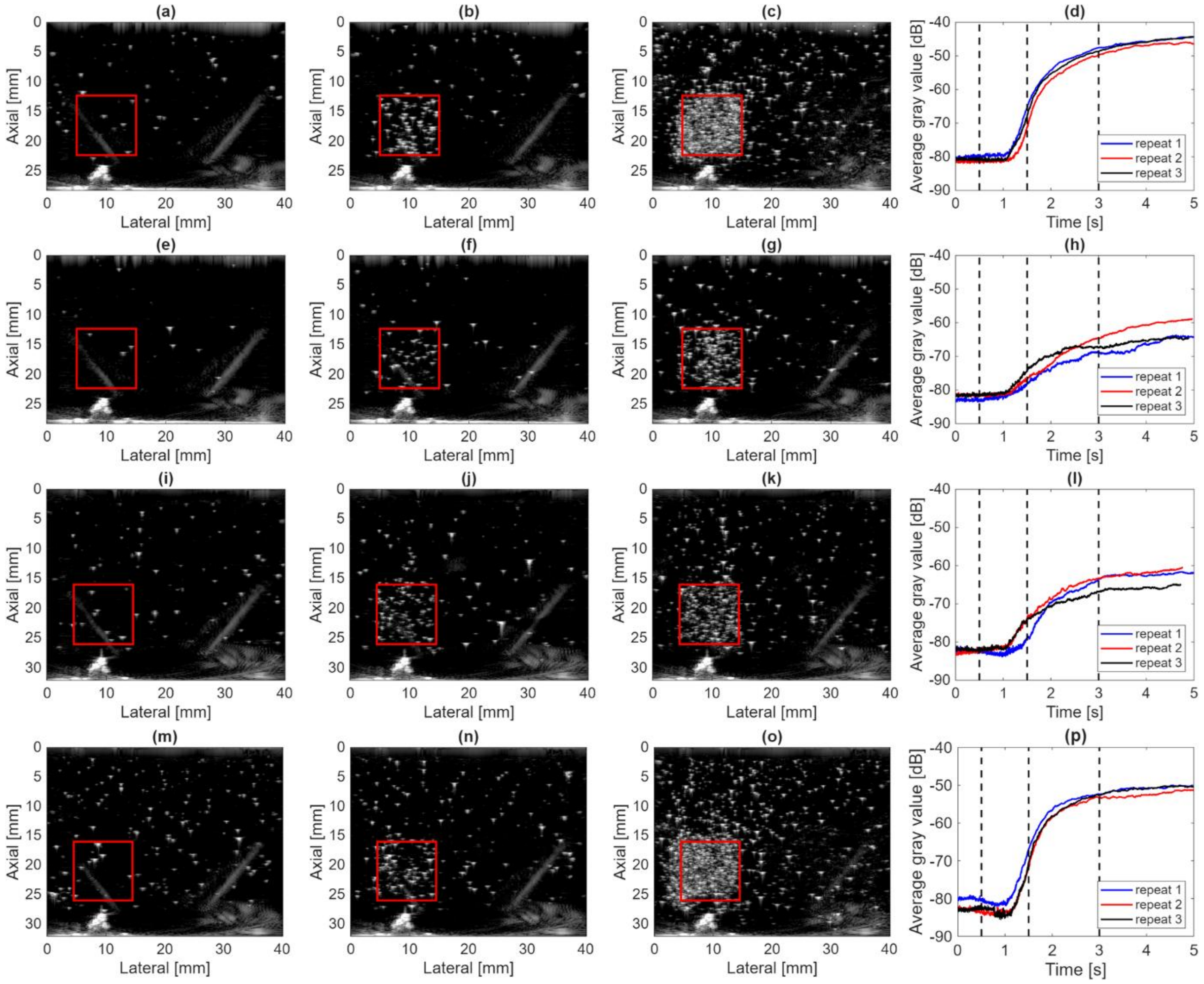

**Figure 4.** Photon radiation response of 17:4 OFP:DFB nanodroplets at 37 °C. B-mode images at different time points with respect to the start of the irradiation: -0.5 s (a, e, i, m), +0.5 s (b, f, j, n) and +2 s (c, g, k, o), shown for a representative phantom of four different nanodroplet vials (a-c, e-g, i-k and m-o). The phantoms were irradiated with a dose of 0.1 Gy at a dose rate of 1.67 Gy/min. The red box indicates the 10 x 10 $mm^2$ radiation field and the quantification ROI. The final column (d, h, l and p) shows the corresponding dynamics of the quantification of the average gray value over time for three different phantoms (prepared from the same vial). The red curve corresponds to the displayed B-mode images. Vertical dashed lines correspond to the different time points of the B-mode images.

A clear radiation response was observed in each case, primarily confined to the radiation field, and with very limited background bubbles present prior to irradiation. Looking at the dynamics of the contrast generation, good reproducibility was observed for phantoms prepared from the same nanodroplet vial. However, a non-negligible variation in both the overall radiation response and the response dynamics was observed for phantoms prepared from different nanodroplet vials, indicating substantial inter-vial variability. For example, while the background signals for all phantoms hovered around -80 dB, the peak signals (plateaus) varied between circa -65 to -50 dB. The variability of the radiation response is also clear from comparing the B-mode images at the final time point. Where in some phantoms the microbubble cloud fills the entire ROI (figure 4c,o), in certain phantoms large empty zones remain (figure 4g,k).

### 3.4 The manual nanodroplet synthesis and different fugacity of the mixed-core liquids result in substantial variability in the radiation response

Two hypotheses could explain the observed inter-vial variability for mixed-core droplets. First, as the photon response strongly relies on the exact degree of superheat of the nanodroplet core, it also strongly depends on its exact composition. To evaluate whether both perfluorocarbons were indeed incorporated in a 17:4 ratio as expected from the synthesis protocol, $^{19}$F NMR spectroscopy was performed (figure 5). While both perfluorocarbons were incorporated in the nanodroplet core, substantial differences in their relative proportion existed between different nanodroplet vials. Moreover, a rapid decrease in OFP:DFB ratio was observed over time, which can further alter the degree of superheat of the nanodroplet core. The impact of the latter on our radiation experiments was minimized by keeping the vials on ice and using the nanodroplet vials immediately after preparation (within ~15 min). The rapid diffusion of OFP out of the droplet core (explaining lower OFP:DFB ratios at later time points), in combination with the 30 min time window between the end of droplet preparation and the end of the first NMR measurement, also explains why even at t = 0 h, the molar ratio 5.3:1 expected from the 17:4 volumetric ratio (dashed horizontal line in figure 5c), is not reached. The inter-vial variability in OFP:DFB ratio at the initial time point could contribute to the variability observed in the radiation response for different vials shown in figure 4.

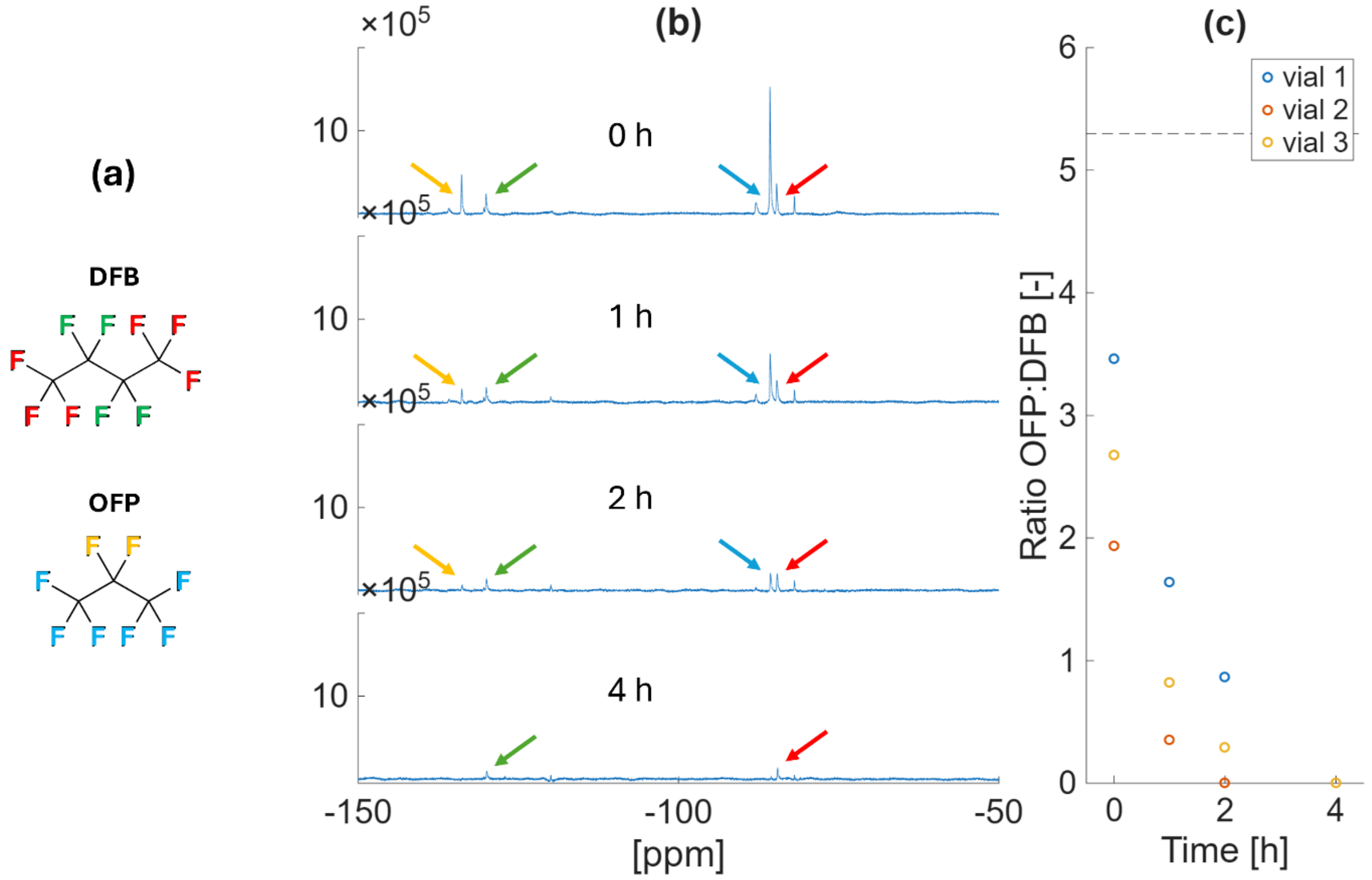

**Figure 5.** [19]F NMR spectroscopy of 17:4 OFP:DFB mixed-core nanodroplets over time. (a) Chemical structure of decafluorobutane (DFB) and octafluoropropane (OFP) with color coding of chemically equivalent fluorine atoms. (b) Evolution of the NMR spectrum over time for a representative nanodroplet sample. Spectra at different time points were scaled using the internal reference at -168 ppm (not shown). (c) Molar ratio of OFP to DFB in the nanodroplet core over time for three independent nanodroplet vials. Horizontal dashed line indicates the expected molar ratio (5.3:1) for a volumetric ratio of 17:4. Vial 1 corresponds to the displayed spectra. At t = 4 h, OFP was no longer detectable, hence all data points overlap at a ratio of 0.

If the variability of the photon response in mixed-core droplets is solely due to the presence of multiple liquids in the nanodroplet core, one would expect no (or much less) variability for pure core droplets. Therefore, we irradiated three phantoms containing pure OFP droplets from different nanodroplet vials at 28 °C. Figure 6 depicts a comparison of the dynamics of the contrast response, revealing the intra- (a) and inter-vial (b) variability in case of pure OFP nanodroplets. Despite the pure core and the limited intra-vial variability, the inter-vial variability remained, as reflected by the variable contrast increase (difference between pre- and post-irradiation plateau).

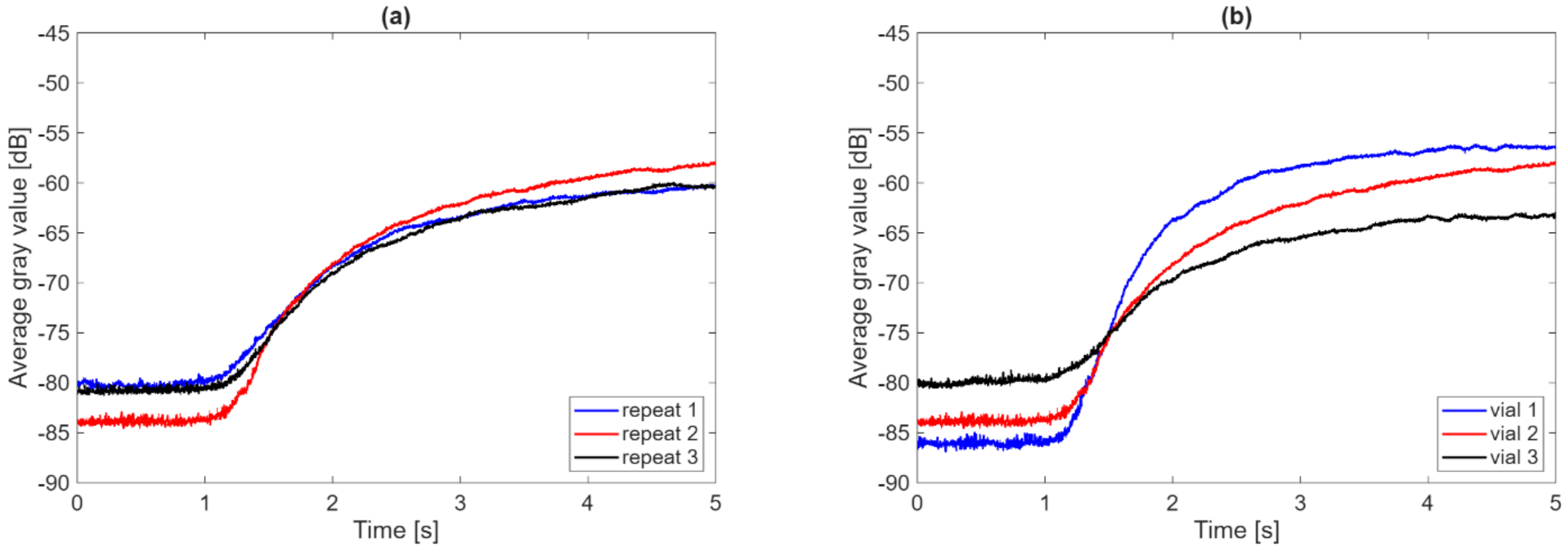

**Figure 6.** Intra- vs inter-vial variability in photon radiation response for pure OFP droplets at 28 °C. Dynamics of the quantification of the average gray value in a 10 x 10 $mm^2$ ROI covering the radiation field (as indicated in figure 2) for three different repeats prepared from the same vial (a) and for three measurements from different nanodroplet vials (b). Phantoms were irradiated with a dose of 0.1 Gy at a dose rate of 1.67 Gy/min. Irradiation starts at ~1 s.

This suggests a second hypothesis, namely that a variation in the initial nanodroplet concentration causes a variability in the radiation response. This can again be evaluated using $^{19}F$ NMR spectroscopy by comparing the initial concentrations of the perfluorocarbons. The results in figure 7 indeed reveal that the perfluorocarbon concentration features substantial differences between different nanodroplet vials. Partially, these differences can be attributed to the variability in the microbubble precursor solution, supported by the Coulter counter measurements shown in figure S3. In addition, part of the sample is inevitably lost in the manual transfer of the microbubble solution to the Luer lock syringe and subsequent condensation.

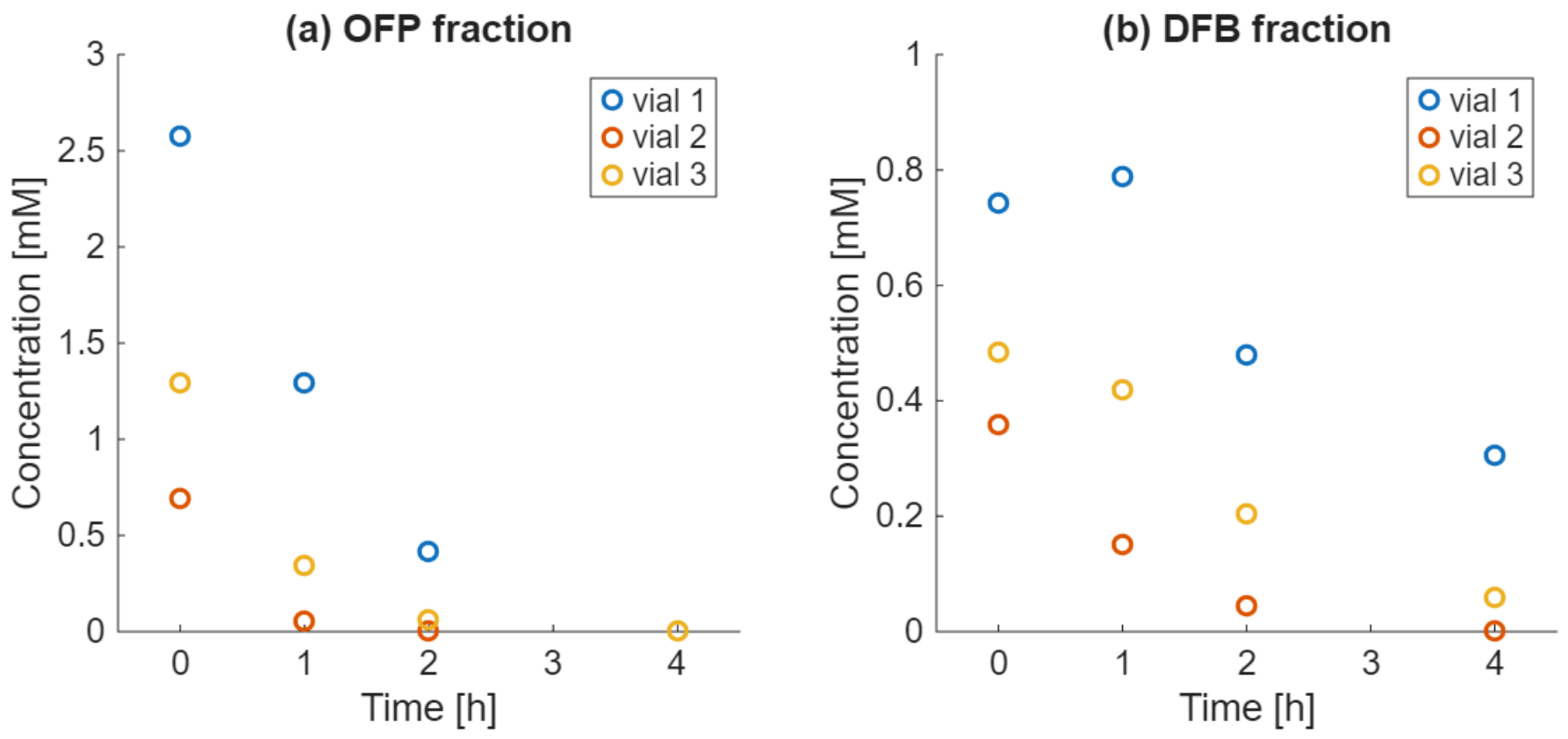

**Figure 7.** Nanodroplet yield over time after microbubble condensation determined via $^{19}F$ NMR spectroscopy. (a) Concentration of the OFP fraction in 17:4 OFP:DFB droplets. OFP was no longer detectable after 4 h. (b) Concentration of the DFB fraction in 17:4 OFP:DFB droplets. Samples were all prepared from 1 mL of microbubble precursor solution.

### 3.5 At clinically-relevant radiation doses, one can modulate the photon response by varying the nanodroplet concentration

To confirm that the amount of radiation-induced ultrasound contrast depends on the nanodroplet concentration and to illustrate that the technology also works for clinically-relevant radiation doses, we irradiated phantoms containing different volumes of nanodroplet sample with 2 Gy of photons. Given the extended duration of the irradiation, free-floating microbubbles had time to flow out of the radiation field. Hence, we only show the response dynamics for different nanodroplet vials (figure 8). A full video of the radiation response of a representative phantom is available as supplementary material (Video S2). Similarly to the short 0.1 Gy irradiations, the signal rapidly reaches a plateau value, which is relatively maintained for the entire irradiation. Different nanodroplet concentrations (200 and 50 μL) yielded different plateau values, with the higher concentration reaching higher ultrasound gray values (figure 8a).

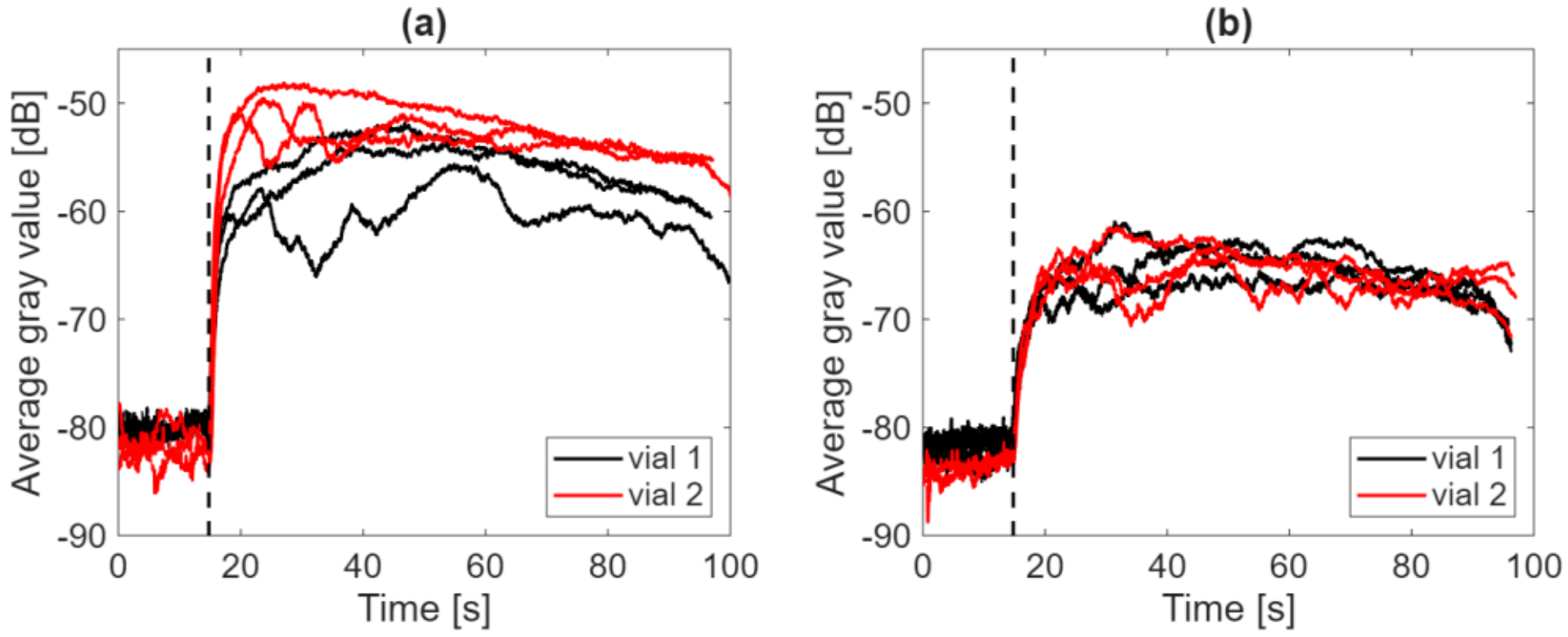

**Figure 8.** Photon radiation response of 17:4 OFP:DFB nanodroplets at 37 °C. Quantification of the average gray value in a 10 x 10 $mm^2$ ROI covering the radiation field for phantoms from different nanodroplet vials and for different nanodroplet volumes: 200 µL (a) and 50 µL (b). The phantoms were irradiated with a dose of 2 Gy at a dose rate of 1.67 Gy/min. Different color curves represent three different repeats prepared from the same nanodroplet vial. Vertical dashed lines indicate the start of the irradiation. Note that for the black curve with the lowest "plateau" in panel (a) we only reached 35.1 °C during irradiation.

## 4. Discussion

We evaluated different photon-sensitive nanodroplet compositions for ultrasound-based photon dosimetry at body temperature. This required a nanodroplet formulation exhibiting a strong photon response on the one hand, while maintaining sufficient stability on the other hand.

### 4.1 Importance of the nanodroplet degree of superheat on the photon radiation response

To illustrate the importance of the degree of superheat on the photon response, we first irradiated OFP droplets at different temperatures. D'Errico previously described a fluid-dependent temperature range in which superheated emulsions are photon-sensitive (between s = 0.51 and s = 0.65, D'Errico, 2001). The vertical dashed lines in figure 9a delimit this photon-sensitive zone for pure OFP. Interestingly, when plotting the average signal increase between the start and end of the irradiation, a linear relationship was observed with the lowest (highest) radiation response observed at the temperature closest to the lower (upper) limit of the photon-sensitive zone. As the dB scale log-compresses the data, this linear behavior is in agreement with the exponential increase in count rates observed by D'Errico with increasing temperature (D'Errico 2001). Unfortunately, the upper limit for OFP is lower than body temperature, preventing its use for in vivo applications. While this upper limit can be shifted to higher temperatures by improving droplet stability via an appropriate shell (e.g. lipid, polymer) or decreasing the nanodroplet size (additional contribution of the Laplace pressure will increase the boiling point, reducing the degree of superheat), the lipid-shelled OFP droplets used in the present study did not survive the experimental procedures to evaluate their radiation response at 37 °C. On the contrary, pure DFB droplets did not exhibit sufficient superheat to be sensitive to photons at 37 °C.

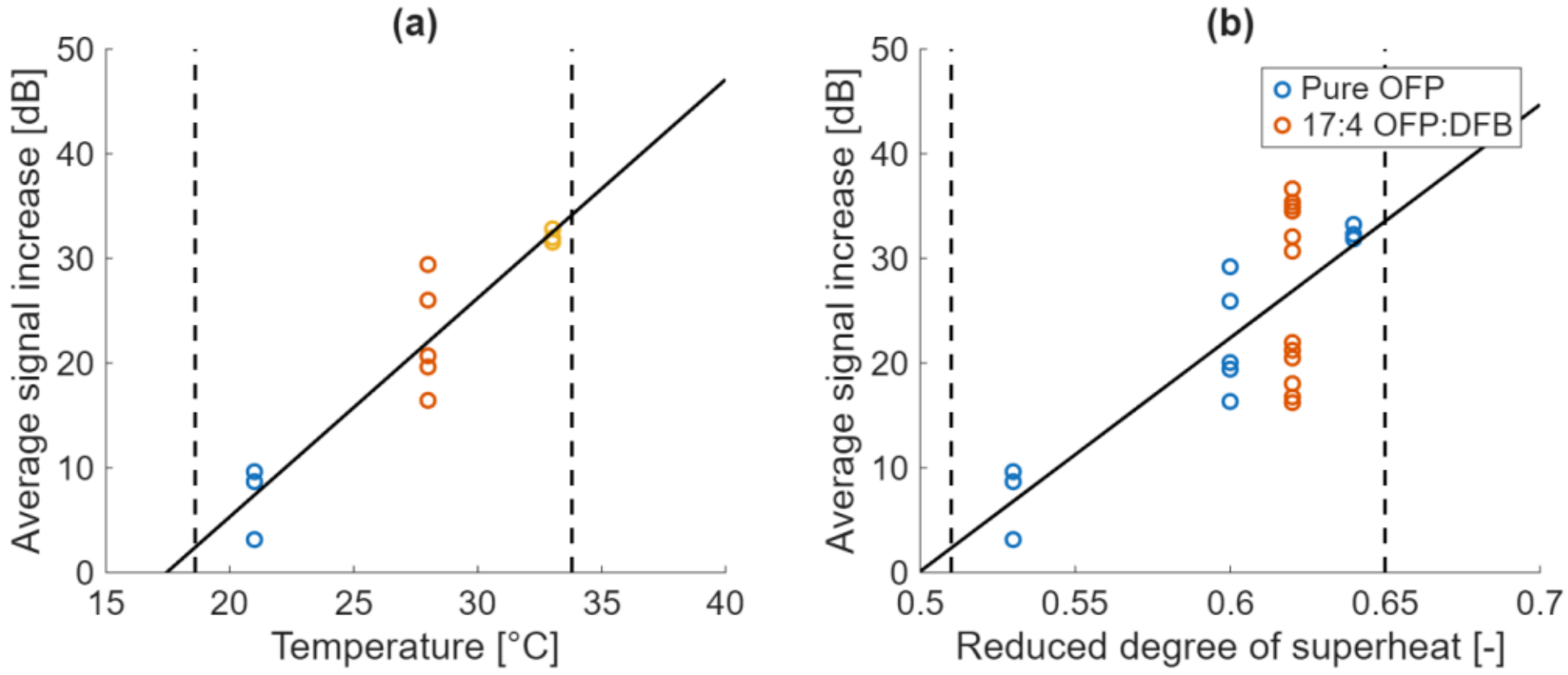

**Figure 9.** Average gray value increase between the start and end of photon irradiation in a 10 x 10 $mm^2$ ROI covering the entire radiation field for pure OFP droplets at varying temperatures (a), and for different droplet compositions at varying degree of superheat (b). Phantoms were irradiated with 0.1 Gy at 1.67 Gy/min. Data points in panel (a) refer to the irradiations presented in figure 2 and figure 6, while in panel (b) also the data points of figure 4 are added. The solid black line shows the linear fit to the data points: (a) y = 2.0890x – 36.4517, $R^2$ = 0.8762; (b) y = 222.9x – 111.3, $R^2$ = 0.5739, while the vertical dashed lines delimit the theoretical photon-sensitive zone of superheated emulsions.

D'Errico also described that the radiation response of different emulsions can be directly compared when the absolute temperatures are converted to the reduced degree of superheat ($s$, equation 1) (D'Errico 2001). In that case, one would expect the radiation response of mixed-core droplets to follow the same trend. While true on average (figure 9b), the large inter-vial variability, reflected in the large drop in $R^2$ value (0.88 to 0.57) of the trendline fit when adding the mixed-core data points, prevents to make conclusive claims. Note that evaluating the radiation response based on the gray value increase is not optimal as the gray values are influenced by saturation of the ultrasound signal, attenuation by the existing microbubble cloud and nanodroplet depletion effects. Nevertheless, figure 9 confirms the importance of the degree of superheat of the nanodroplet core to tune the photon response.

### 4.2 Importance of the nanodroplet formulation on the photon radiation response

Based on theoretical considerations, we opted for a 17:4 OFP:DFB nanodroplet formulation corresponding to a reduced degree of superheat $s \approx 0.63$. This formulation yielded a strong photon response with minimal background bubbles due to spontaneous vaporization. Nevertheless, the strong fugacity (= tendency to escape the core (Brambila *et al* 2020)) of OFP in combination with the thin lipid shell and manual synthesis method, resulted in a short shelf-life with strong fluctuations in the nanodroplet core composition and overall yield. These fluctuations manifested themselves in substantial inter-vial variability, which has to be overcome to enable quantitative dose read-outs.

Lipid-shelled nanodroplets (and microbubbles) are notorious for their poor shelf-life. Sometimes thicker, polymeric shells are preferred (Hadinoto *et al* 2013, Sirsi and Borden 2009). While they reduce diffusion out of the core, they are unfortunately not compatible with the microbubble condensation method. Indeed, lipid shells can reorganize by shedding excess lipids upon volume compression. Instead, rigid polymeric shells tend to break, generating local openings through which the core rapidly escapes (Bloch *et al* 2004). Hence, to prepare polymeric nanodroplets, the core material needs to be liquid from the start of the synthesis. This raises challenges for the used low-boiling point liquids. While OFP (boiling point = -37 °C) can be easily liquified e.g. by placing a glass vial in liquid nitrogen, the aqueous lipid solution would instantly freeze, preventing droplet formation. Similarly, microfluidic approaches have been proposed to improve the reproducibility of the nanodroplet production (Martz *et al* 2012, Melich *et al* 2020), but they feature the same challenges for low-boiling point liquids. In addition, they suffer from low production yields. An alternative synthesis process called the ouzo method, relies on saturating a co-solvent with perfluorocarbons before adding a poor solvent with dissolved lipidic shell precursors (Li *et al* 2019). This forces droplet nucleation, even for gaseous low-boiling point perfluorocarbons. Yet, the monolayer lipid shell still limits long-term droplet stability. While some flexibility exists in modifying the lipid composition to modulate shell permeability (Garg *et al* 2013), more promising is the transition towards hybrid lipid-polymer shells (Hadinoto *et al* 2013). Recently, Wang and colleagues showed that adding pluronic F-68 to a lipid microbubble shell strongly enhances the microbubble stability by preventing coalescence (Wang *et al* 2025). As part of our future work, we aim to explore whether such enhanced microbubble stability also translates to an improved nanodroplet shelf-life.

If an equally responsive, but more stable nanodroplet formulation can be found, most of the limitations discussed above are easily solved. Firstly, reduction of diffusion out of the nanodroplet core would prevent rapidly fluctuating core ratios in mixed-core droplets, ensuring a more stable radiation response. Secondly, if shelf-life is improved, nanodroplet concentrations can be determined prior to irradiation (e.g. via nanoparticle tracking analysis), to account for residual differences in nanodroplet yield, further reducing inter-vial variability. Thirdly, a centrifugal size selection step could be added to the protocol after microbubble condensation to yield a more monodisperse nanodroplet population (Collado-Lara *et al* 2023), again minimizing the variability of the radiation response.

In this paper, we only explored a single mixture of perfluorocarbons. Other mixtures might exhibit a more optimal radiation response. While optimizing the mixture is important, this is best performed when testing the droplets in an in vivo context. Intravenous injection, blood pressure fluctuations, (bio)chemical interactions, etc. may all influence the nanodroplet stability in vivo. Depending on the exact requirements these factors impose on the droplet stability and thus the upper limit of superheat that can be used in an in vivo setting, the droplet composition can be further improved.

### 4.3 Limitations and future perspectives

To minimally influence the photon response, we opted for freely floating nanodroplets, which were not stirred during irradiation. This caused the generated microbubbles to flow out of the irradiated zone over time as particularly noticeable for the long irradiations (video S2). This also limited the replenishment of nanodroplets to passive diffusion. Remarkably, signal plateaus were largely maintained for the entire 2 Gy irradiation, indicating that the outflow of microbubbles was balanced by the vaporization of fresh nanodroplets. Moreover, different plateau levels were reached for different nanodroplet concentrations, indicating the possibility to modulate the radiation-induced ultrasound response by varying the nanodroplet concentration. This is in agreement with earlier findings that the radiation response is dependent on the nanodroplet concentration (volume fraction) and the effective fluence of particles contributing to vaporization events (in this case photon-induced electrons) (Collado-Lara *et al* 2023, Carlier *et al* 2026). In vivo, we envision to detect vaporization events directly in circulation. Depending on the circulatory lifetime of the droplets, we expect to continuously replenish the nanodroplet supply and wash out generated microbubbles, preventing rapid saturation of the ultrasound signals. Noteworthy, in vivo, nanodroplets would be confined by vessels, potentially influencing their vaporization response as previously observed for acoustic droplet vaporization (Lin *et al* 2017). While the radiation stimulus will be less affected by attenuation than an ultrasound trigger, the limited room for expansion may reduce the number of observed vaporization events.

Quantification of the (dynamics of) the radiation-induced ultrasound contrast relied on the evaluation of gray values extracted from conventional B-mode sequences. While sufficient for simple phantom setups with minimal background, more adequate contrast sequences will be required in vivo. For this purpose, we are currently implementing dedicated amplitude modulation (AM) and amplitude-modulated pulse inversion (AMPI) sequences, which are particularly sensitive to the non-linear oscillations of generated microbubbles and "mute" tissue background (Mine *et al* 2024). In addition, we are developing an ultrasound localization microscopy (ULM) pipeline (Errico *et al* 2015, Collado-Lara *et al* 2022, Heymans *et al* 2025). Benefitting from high-frame rate acquisitions or passive detection, we should be able to localize individual vaporization events with µm-level precision. By tracking bubbles over subsequent frames, one can (partially) discriminate between newly generated bubbles and bubbles flowing around. However, using 2D ultrasound imaging, it remains difficult to distinguish between droplets flowing through the elevational plane and those that are newly vaporized. To overcome this problem, we are setting up a matrix probe eventually enabling to capture vaporization events and track generated microbubbles in 3D.

Reliable vaporization counts would enable to establish truly quantitative dose-response relationships, if the nanodroplet concentration and radiation dose in the blood stream are known. The former can be determined by evaluating the circulatory lifetime of the contrast agent and by choosing the appropriate injection approach (bolus vs continuous infusion), while the latter would be simulated using the HEDOS framework (Shin *et al* 2021).

## 5. Conclusion

For the first time, we demonstrated strong photon-induced ultrasound contrast using mixed-core nanodroplets, paving the way for future in vivo dosimetry applications. We illustrated the importance of the degree of superheat, with nanodroplets exhibiting a photon response only within a relatively small thermal range. A trade-off exists between a stronger photon response (higher degree of superheat) and the stability of the droplet formulation. In this regard, pure perfluorocarbon cores offer insufficient flexibility to modulate the response at body temperature as illustrated for octafluoropropane (OFP, spontaneous vaporization) and decafluorobutane (DFB, no photon response) cores. Mixed-core droplets allowed to tune the degree of superheat to ensure a strong radiation response at body temperature, as demonstrated for a 17:4 OFP:DFB core for photon doses as small as 0.1 Gy. The use of perfluorocarbons with different fugacity and the employed synthesis approach resulted in substantial variability in the nanodroplet yield and composition. Core composition also rapidly changed over time due to the preferential diffusion of the more volatile (OFP) compound. These factors manifested themselves in non-negligible inter-vial variability of the radiation response. Alternative nanodroplet formulations with an improved stabilizing shell will be required to strengthen reproducibility. In combination with the transition from semi-quantitative gray value read-outs to the detection of individual vaporization events, this may eventually enable quantitative dosimetric read-outs.

## 6. Acknowledgments

The authors want to thank Hilde Geeraerts for the execution of the phantom irradiations and Fatemeh Mashayekhi for assistance with the ultrasound acquisitions. The project received funding from the Research Foundation Flanders (FWO) under grant agreement No. G044425N. UH received funding under FWO grant agreement No. G057825N. SVH received a postdoctoral fellowship from the Research Foundation Flanders (No. 12E0923N).

## Supplementary materials

Below the supplementary figures and video captions referred to in the manuscript: 'Towards in vivo photon dosimetry using radiation-sensitive mixed-core ultrasound phase-change contrast agents' are listed with their corresponding description.

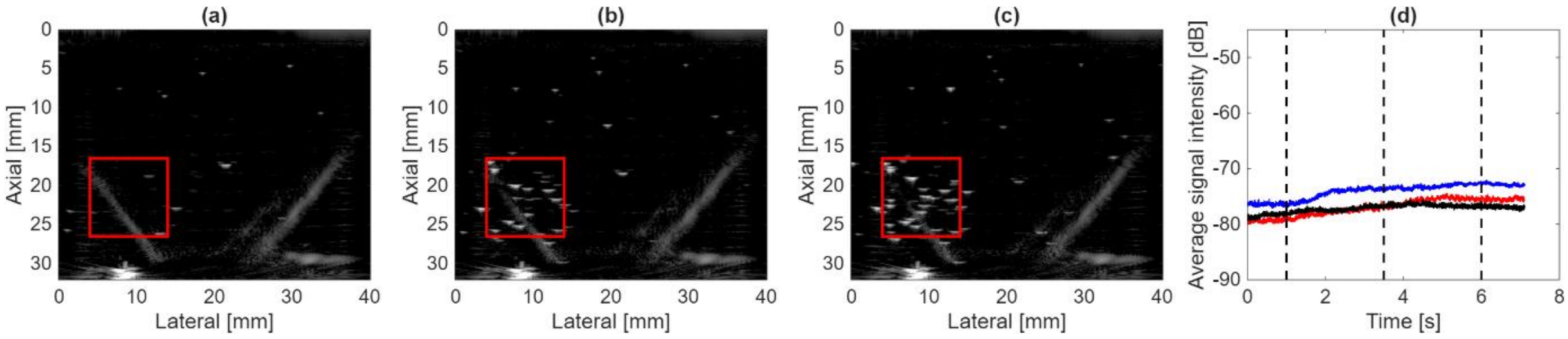

**Figure S1.** Photon radiation response of pure OFP nanodroplets at 21 °C. B-mode images at different time points during the ultrasound acquisition: 1 s (a), 3.5 s (b), and 6 s (c). The phantoms were irradiated with a dose of 0.1 Gy at a dose rate of 1.67 Gy/min (start of irradiation could not be determined from the contrast increase). The red box indicates the 10 x 10 $mm^2$ radiation field. The final subfigure (column d) shows the corresponding dynamics of the quantification of the average gray value in the red ROI for the entire ultrasound acquisition for three different phantoms (prepared from the same vial). The red curve corresponds to the displayed B-mode images. Vertical dashed lines correspond to the different time points of the B-mode images.

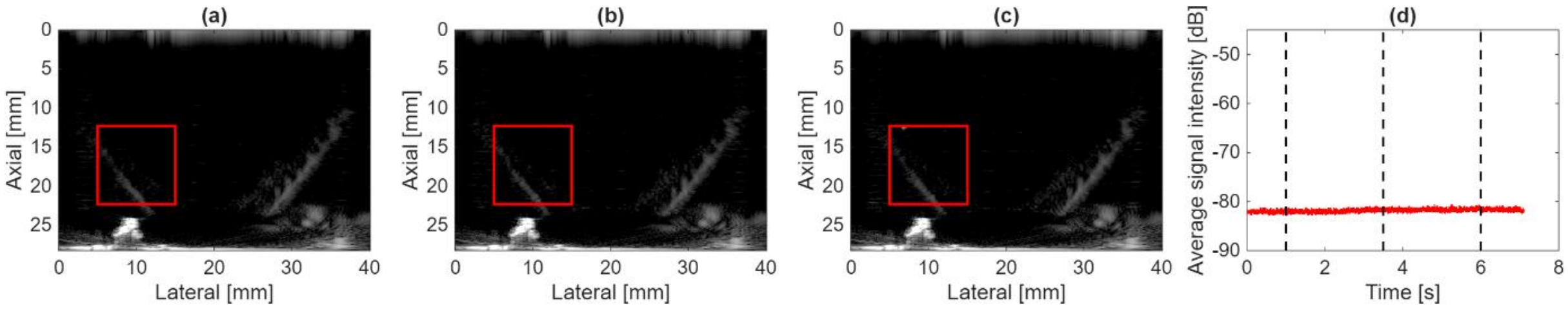

**Figure S2.** Photon radiation response of a water phantom without nanodroplets. B-mode images at different time points during the ultrasound acquisition: 1 s (a), 3.5 s (b), and 6 s (c). The phantom was irradiated with a dose of 0.1 Gy at a dose rate of 1.67 Gy/min (start of irradiation at ~3 s). The red box indicates the 10 x 10 $mm^2$ radiation field and the quantification ROI. The final subfigure (column d) shows the corresponding dynamics of the quantification of the average gray value for the entire ultrasound acquisition. Vertical dashed lines correspond to the different time points of the B-mode images.

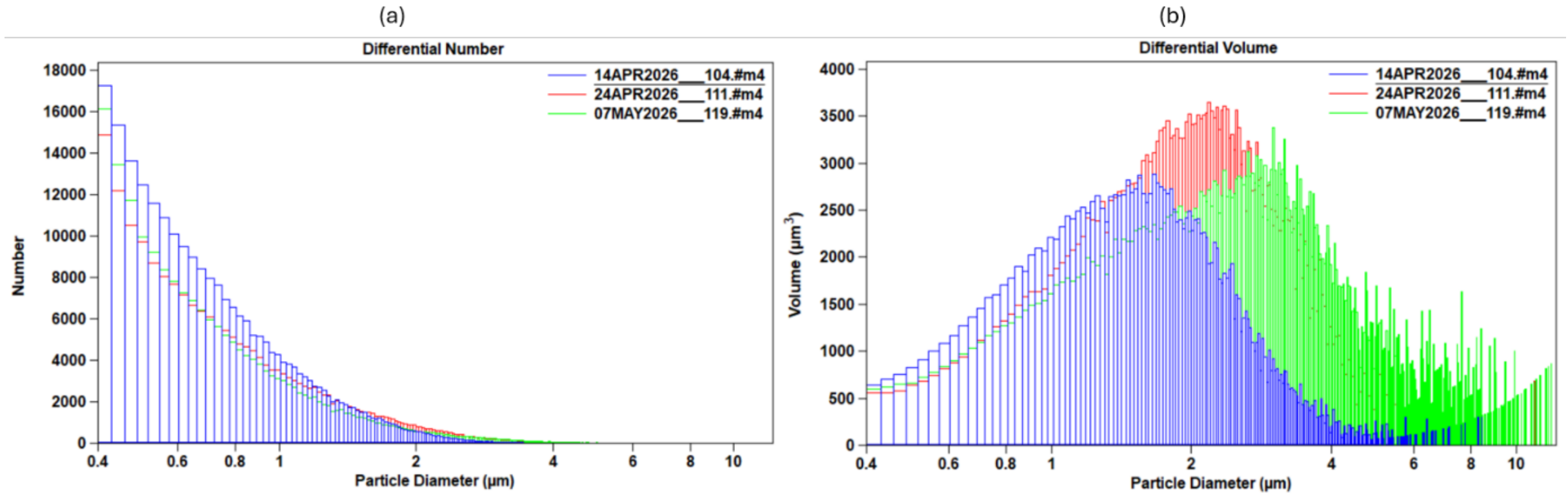

**Figure S3.** Microbubble yield and size distribution prior to condensation as determined through Coulter counter measurement, plotted according to number (a) and volume (b). Different colors indicate different vials of 17:4 OFP:DFB microbubbles measured on different days.

**Attached videos**

**Video S1.** Photon radiation response of 17:4 OFP:DFB nanodroplets (200 μL) at 37 °C. Online B-mode images (a) and quantification of the average gray value in the red ROI (b) for the entire ultrasound acquisition at a 400 Hz frame rate. The phantom was irradiated with a 0.1 Gy dose at a dose rate of 1.67 Gy/min. The irradiation starts at ~2s.

**Video S2.** Photon radiation response of 17:4 OFP:DFB nanodroplets (200 μL) at 37 °C. Online B-mode images (a) and quantification of the average gray value in the red ROI (b) for the entire ultrasound acquisition at a 27 Hz frame rate. The phantom was irradiated with a 2 Gy dose at a dose rate of 1.67 Gy/min. The irradiation starts at ~21s.